# Quantum entanglement enabled ellipsometer for phase retardance measurement

Meng-Yu Xie, [a,b,1]　Su-Jian Niu, [a,b,1]　Yin-Hai Li, [a,b,1]　Zheng Ge, [a,b]　Ming-Yuan Gao, [a,b] Zhao-Qi-Zhi Han, [a,b]　Ren-Hui Chen, [a,b]　Zhi-Yuan Zhou, [a,b,*]　and Bao-Sen Shi [a,b,*]

　　a CAS Key Laboratory of Quantum Information, University of Science and Technology of China, Hefei, Anhui 230026, China

　　b CAS Center for Excellence in Quantum Information and Quantum Physics, University of Science and Technology of China, Hefei 230026, China

　　1 These three authors contribute equally to this article

　　* Corresponding authors. Email: drshi@ustc.edu.cn; zyzhouphy@ustc.edu.cn

## Abstract

An ellipsometer is a vital precision tool used for measuring optical parameters with wide applications in many fields, including accurate measurements in film thickness, optical constants, structural profiles, etc. However, the precise measurement of photosensitive materials meets huge obstacles because of the excessive input photons, therefore the requirement of enhancing detection accuracy under low incident light intensity is an essential topic in the precision measurement. In this work, by combining a polarization-entangled photon source with a classical transmission-type ellipsometer, the quantum ellipsometer with the PSA (Polarizer-Sample-Analyzer) and the Senarmount method is constructed firstly to measure the phase retardation of the birefringent materials. The experimental results show that the accuracy can reach to nanometer scale at extremely low input intensity, and the stability are within 1% for all specimens tested with a compensator involved. Our work paves the way for precision measurement at low incident light intensity, with potential applications in measuring photosensitive materials, active-biological samples and other remote monitoring scenarios.

## 1. Introduction

Elliptic polarization measurement (also called ellipsometry) is a technology that accurately measures various physical properties by detecting the polarization changes before and after the sample [1, 2], including optical constants, film thickness, lattice vibration modes, local atomic structure [3] and so on. The measurement accuracy of the ellipsometry is closely related to the intensity of the pump laser. In general, the theoretical measurement is limited by the standard quantum limit of $1/\sqrt{N}$, and thus the error can be reduced by increasing the number of detected photons $N$. However, in practical applications, the characteristics of a material could be impacted by the strong pump laser. For example, photosensitive materials exposed to strong light may produce photochromic, polymerization, fracture, deformation, and other changes in physical properties, and the light intensity reaching the damage threshold of the material may cause irreversible photo-thermal damage and others. Hence, enhancing detection accuracy under low incident light intensity is an essential topic in the field of precision measurement.

The entangled photon source based on the unique quantum feature provides a feasible method to accurately measure the optical properties of a material by using ultralow light intensity illumination. Specifically, in the field of quantum metrology, an *N*-particle entangled source could improve the theoretical measurement limit by a factor of $1/\sqrt{N}$, from the standard quantum limit to the Heisenberg limit [4-6]. Thus far, by utilizing the entangled photon source generated by spontaneous parametric down conversion in a nonlinear crystal, the precise measurements exceeding the standard quantum limit have been achieved [7-10]. At the beginning of the 21st century, Teich *etc*. proposed the basic configuration of an ellipsometer based on non-classical photon sources in 2001[11, 12], and showed the experimental test using correlated photon in 2004 [13]. So far, theoretical works mainly concentrate on the establishment of a quantum ellipsometry framework by analogizing quantum expressions based on traditional systems, for example, analogizing the density matrix as the coherence matrix [14] and analogizing the quantum completely positive map as the Mueller matrix [15]. Meanwhile, in experiments, studies have measured the concentration of chiral solutions [16-18] and image the biological samples [19, 20] by using the entangled photon source.

All above works are intrinsically based on detecting the variance of the polarization states. So far, there is no any work reporting the measurement of the birefringence of a material with entangled photon source. In addition, given that most of works are confined to the basic PSA (Polaizer-Sample-Analyzer) structure, further enhancement on the measurement accuracy could be achieved using some compensation methods. In this work, we derive a general theoretical description of quantum ellipsometry for measuring the retardance of birefringent materials, and show the experimental proof using a quantum-enabled dual-way structure. Furthermore, to improve the accuracy of measurement, the classical Senarmount method is introduced firstly to the quantum stystem. The experimental results show that the long-term deviations from the standard values are within 1% for all specimens tested with a compensator involved, the accuracy can reach the nanometer scale at extremely low intensity, which is superior to the classical counterpart at the same input level, and equivalent to the classical Senarmont method using conventional laser sources. Our work shows that the quantum ellipsometry may have some potential applications in measuring photosensitive materials, active-biological samples and other remote monitoring scenarios.

## 2. Methods
### 2.1 Theoretical models

The ellipsometer measures the polarization of the probe light before and after the target sample. Analogous to the classical transmission-type ellipsometer, we use the entangled light source to build a quantum ellipsometer, aiming to measure the birefringence of a sample precisely. The quantum ellipsometry system is devised based on the traditional PSA, and the flow chart of the system is shown in Fig. 1(a). More comparisons between classical and quantum ellipsometers can be found in the appendix A.

The polarization-entangled photon source is used as the light source of the transmission-type ellipsometry system, which is separated by a dual-way structure named signal and idler paths. Suppose that the state of the input polarization-entangled photons is chosen to be one of the four Bell states, described as $|\phi^+\rangle = \frac{1}{\sqrt{2}}(|HV\rangle + |VH\rangle)$. The sample is placed in the idler path while the signal path remains nothing before entering the analyzer. Here, we regard $J_{s(i)}$ as the evolution operator of photons in the signal(idler) path, therefore the two-photon state after passing through the sample is

$$|\phi_{sample}^+\rangle = J_s \otimes J_i |\phi^+\rangle = \frac{1}{\sqrt{2}}(A|HH\rangle + B|HV\rangle + C|VH\rangle + D|VV\rangle). \tag{1}$$

Here we have $J_s = I$ and $J_i = S$, with I denoting the unit matrix and S standing for the Jones matrix of the sample. The probability amplitudes are $A = -\sin\theta\cos\theta(e^{i\delta} - 1)$, $B = \sin^2\theta + e^{i\delta}\cos^2\theta$, $C = \cos^2\theta + e^{i\delta}\sin^2\theta$ and $D = -\sin\theta\cos\theta(e^{i\delta} - 1)$. $\theta$ marks the angle of the slow axis of the birefringent material, and $\delta = \frac{2\pi}{\lambda}|n_e - n_o|d$ shows the phase retardation between horizontally and vertically polarized light after passing through the sample with $d$ representing the sample thickness.

From the above expression, the phase retardation $\delta$ is already encoded in the probability amplitude of the entangled state. The HWP and PBS are employed here to perform the projection measurement. By rotating the HWP on either the signal or idler side, we can select a joint measurement base that is suitable for the sample birefringence measurement. Therefore, the coincidence measurements under arbitrary joint projective measurement (for details, see appendix A) can be represented as

$$I_{out1} = \frac{I_0}{4}[2 - (1 + \cos\delta)\cos(4(h_s + h_i)) - (1 - \cos\delta)\cos(4(h_i - h_s - \theta)],$$
$$\frac{\partial I_{out1}}{\partial \delta} = \frac{I_0}{4}\sin\delta[\cos(4h_i + 4h_s) - \cos(4h_i - 4h_s - 4\theta)]. \tag{2}$$

Here, $I_0$ represents the initial photon coincidence counts, and $h_{s(i)}$ is the HWP angle in the signal(idler) path. The phase retardation $\delta$ of the sample can be obtained by fitting the intensity curve. If we fix the measurement base and the angle of the sample ($h_s$ and $\theta$ are fixed), it can be seen from Eq. (2) that the output coincidence photon counts are the functions of the analyzer's angle $h_i$. Therefore, the fit curve between Eq. (2) and output photon counts shows the phase retardation and the initial coincidence counts in a single sample measurement.

However, in the same situation as traditional PSA, if the phase retardation of the sample is approximately 0 or π, the derivative of the output intensity is close to 0. That is to say, the change in coincidence counts is indistinguishable with a slight variation in phase retardation, thus the measurement accuracy has more severe declines for certain samples. Considering that testing a sample with unknown reference values is common in practice, we need to improve the quantum ellipsometer to measure arbitrary birefringent materials with high accuracy. To solve this problem, we introduce a compensator with the rotating angle fixed at 0° to the idler path (shown as the dotted portion in Fig. 1(b)); therefore, the light intensity and its derivative are calculated as:

$$I_{out2} = \frac{I_0}{4}\left[1 - \cos4h_i\left(\cos4h_s\cos^2\frac{\delta}{2} + \cos(4h_s + 4\theta)\sin^2\frac{\delta}{2}\right) - \sin4h_i\sin(4h_s + 2\theta)\sin\delta\right],$$
$$\frac{\partial I_{out2}}{\partial \delta} = \frac{I_0}{4}\sin(4h_s + 2\theta)(-\cos\delta\sin4h_i + \cos4h_i\sin\delta\sin2\theta). \tag{3}$$

Suppose that an H-basis is chosen for the HWP in the signal path, while the sample axis is set to 45°, then Eq. (3) becomes

$$I_{outx} = I_{out3}|_{h_s=0,\theta=\pi/4} = \frac{I_0}{2}\sin^2\left(\frac{\delta}{2} - 2h_i\right), \quad \frac{\partial I_{outx}}{\partial \delta} = \frac{I_0}{4}\sin(\delta - 4h_i). \tag{4}$$

There are two advantages of this angle combination. First, for an arbitrary phase retardation of the birefringent sample, the output photon counts have noticeable variations, namely max $\max(I_{outx}) - \min(I_{outx}) \equiv I_0/2 > 0$, enough to encounter the environmental noise disturbance, improving the robustness of the quantum ellipsometer. Second, the maximum derivative value is nonzero for arbitrary birefringent samples, which allows a noticeable distinction between the output curves with a slight variation in sample birefringence, thus ensuring the sensitivity of the quantum ellipsometer.

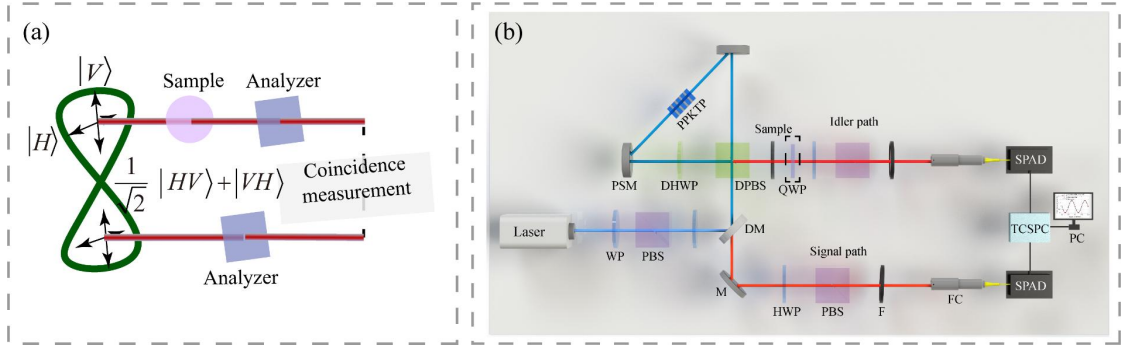

Fig. 1. Theoretical and experimental frameworks of quantum ellipsometer. (a) The flow charts of the quantum ellipsometry. (b) Experimental setup for the quantum ellipsometer based on the Sagnac-loop configuration. WP, wave plate set, including an half-wave plate (HWP) and quarter-wave plate (QWP); PBS, polarizing beam splitter; DM: dichroic mirror; DPBS, dichroic PBS; DHWP, dichroic HWP; PSM, off-axis parabolic silver mirror; PPKTP, periodically poled KTP crystal; F, filter; FC, fiber collimator. (c and d) Characterization of the entangled source. SPAD, Single-photon Avalanche Diode. TCSPC, Time-Correlated Single Photon Counting.

### 2.2 Experimental Setup

A schematic diagram of the quantum ellipsometer is sketched in Fig. 1(b), which consists of three parts: photon pair generation, photon-sample interaction, and polarization analysis. In the photon pair generation part, high quality polarization-entangled photon source is generated via a spontaneous parametric down conversion (SPDC) process in a type-II phase-matched periodically poled $KTiOPO_4$ (PPKTP) crystal (Raicol Crystals Ltd.) embedded in a Sagnac interferometer [21]. A CW laser with a 2mW pump power at 405 nm is derived from a diode laser (Kunteng Quantum Technology Co. Ltd). The pump laser is focused on the center of the PPKTP crystal by two symmetrical parabolic silver mirrors with a focal length of 101.6 mm. The PPKTP crystal has dimensions of 1 mm *2 mm * 20 mm, with a polling period of 10.02 μm, and is fixed in the middle of the Sagnac loop. The photon degenerate wavelength temperature of the PPKTP crystal is set at 23.6 °C with a temperature stability of ± 0.002 °C.

In the photon-sample interaction part, the idler photon passes through the target sample with birefringence information encoded on the evolution of entangled states. The selection of the sample axis can be fairly flexible as long as it is not parallel to the polarization of the idler photon. Here we choose two commercial wave plates, multi-order QWP (808 nm, MFOPT Ltd.) and true zero-order HWP (808 nm, Thorlabs) as samples, whose optical axes are set in various directions between 0 and $2\pi$. The QWP (808 nm, MFOPT Ltd.) in the dotted box is used to improve the accuracy of the measurement, similar to the compensation wave plate in the classic Senarmont method [22].

In the polarization analysis part, both signal and idler photons pass through a true zero-order HWP (808 nm, Thorlabs) for the basis selection and an 808 nm PBS for the projective measurement. The pump beam is removed with a longpass filter (Thorlabs, FELH500 nm) and bandpass filter (Thorlabs, FBH810-10 nm) before collecting into a single-mode fiber. After that, the heralded photons are detected by a single-photon avalanche diode (Excelitas, SPCM-AQRH) with a detection efficiency of approximately 60% at a wavelength of 810 nm. The two-photon states $|\phi_{sample}^+\rangle$ are then reconstructed by the coincidence measurements under different measurement bases, thus, the systematic accuracy and error of the birefringence phase retardation measurement with a certain projective basis can be evaluated accordingly.

## 3. Result and discussion

### 3.1 Characters of the polarization entangled photon source

Before evaluating the performance of the quantum ellipsometer, we test the quality of the polarization entangled photon source. Notably, high fidelity is fairly important in this measurement since the theoretical model is based on a preconfigured maximally entangled state. By removing the sample from the idler path, we measure the two-photon polarization interference curve and perform quantum state tomography to characterize the quality of the generated polarization-entangled state in Fig. 2. The raw interference visibilities ( $V = \frac{C_{max}-C_{min}}{C_{max}+C_{min}}$ ) are calculated as 99.75% ± 0.01% and 99.12% ± 0.01% for H-base (fixed at 0°) and D-base (fixed at 22.5°), respectively, far beyond the 71% shown in the Bell inequality. Moreover, the CHSH inequality to quantify the entanglement quality is measured, and the S-parameter is calculated to be 2.754 ± 0.006191 in 10 s, violating the CHSH inequality by 122 standard deviations [23]. The fidelity between the experimental state $\rho_{exp}$ and the ideal Bell state is estimated to be 99.79% ± 0.013%. The above series of characterizations shows that the current photon source is fairly close to the maximum entangled state, providing a solid basis for making an accurate birefringence measurement of target samples.

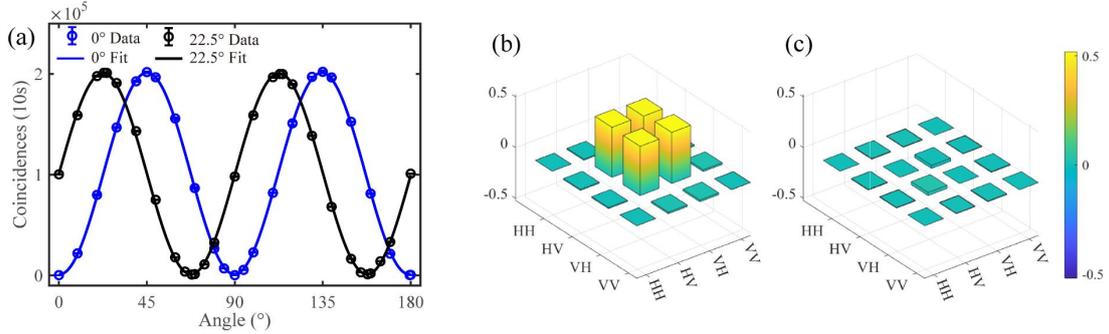

Fig. 2. Properties of the polarization entangled photon source. (a) Coincidence counts in 10 s as a function of the HWP angle with the horizontal (diagonal) projection bases. The background dark coincidence is not subtracted; the data are fitted to the sinusoidal function; error bars are obtained from multiple measurements. (b) and (c) The real and imaginary parts of the reconstructed density matrix of the prepared polarized entangled source using the maximum-likelihood estimation method

### 3.2 Quantum ellipsometer without a compensator

Based on the high quality entangled source, we could move one step further to discuss a quantum ellipsometer system without the QWP compensator in the idler path. Before building the measurement system, the angle of the sample and the waveplates in the polarization analysis part should be determined in advance. The relation between the sample axis $\theta$, the angles of the HWPs $h_s$, $h_i$ and the output photon counts $I_{out}$ is already shown in Eq.(2). Generally, a valid angle combination of sample axes and measurement bases only requires that the output photon counts depend on the phase retardance, in other words, the coefficient before $\cos\delta$ in Eq.(2) should be nonzero:

$$2\sin(4h_i - 2\theta)\sin(4h_s + 2\theta) \neq 0. \qquad (5)$$

In our experiment, we fixed the HWP angle in the signal path while rotating the HWP in the idler path, therefore it requires $4h_s + 2\theta \neq k\pi, k \in \mathbb{Z}$. In fact, this condition ensures that the linear polarization of the input photons has a nonzero angle with either the slow or fast axis of the sample.

Based on the above condition, the coincidence photon counts as a function of the HWP angle $h_i$ can be measured when the sample angle is fixed under an arbitrary joint measurement base. The phase retardation of the sample is obtained by fitting the measured data with the least square method, which sets $I_0$ and $\delta$ both as the fitted values. The results are shown in Fig. 3. In Fig. 3(a), the phase retardation of a true zero-order HWP is measured with two polarization bases, H-base ($h_s$ =0°) and D-base ($h_s$ =22.5°). With the axis of the sample fixed at 22.5°, the two estimated phase retardations in the H-base and D-base are measured to be 3.1417 and 3.1401 rad, respectively. This single measurement shows the plausibility for phase redardance measurement using the quantum ellipsometer. To eliminate the occasional factors, we make multiple measurements in Fig. 3(b) and 3(c), showing the accuracy and stability on various occasions.

In Fig. 3(b), we measure the phase retardation with varying sample axes under H- and D-basis (shown as purple and blue, respectively). The left y-axis shows the phase retardation, while the right y-axis shows the relative errors defined as $\Delta\delta = |\delta_{exp} - \delta_{std}|/\delta_{std}$. The black solid line shown in the middle represents the reference value given by the manufacturer, with a paler shade representing the possible error in the production process. In this way, with a reference value shown as $\delta_{std}$=3.1341 ± 0.0628 rad, the average phase retardations in the H-basis and D-basis are calculated to be 3.0832 ± 0.0703 rad and 3.0810 ± 0.0434 rad respectively, with the majority of the relative errors below 5% in both cases. In addition, Fig. 3(c) has similar organization as Fig. 3(b). It measures the time stability over 2.5 hours under H- and D-basis. The results show that the mean values of the phase retardation are 3.1131 ± 0.0321 rad and 3.1354 ± 0.0110 rad, with the relative errors of -0.67% ± 1.02% and 0.04% ± 0.35%, respectively.

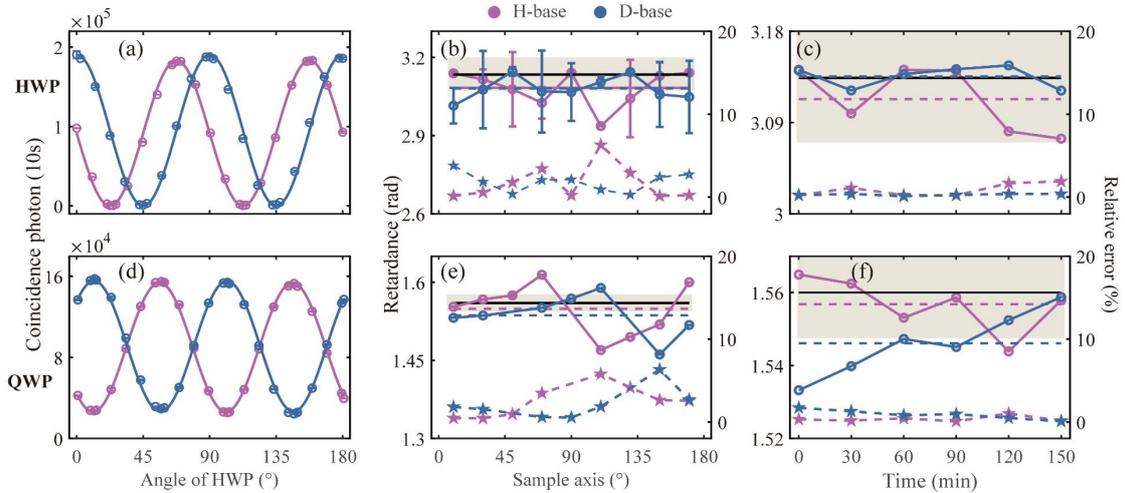

Fig. 3. Phase retardance measurements of the compensation-free quantum ellipsometer. (a) and (d): Single measurement for HWP and QWP. Error bars are obtained from multiple measurements. (b) and (e): Phase retardation under varying sample axes. (c) and (f): The time stability over 2.5 hours.

One problem exposed in the above experiment is that phase retardation is highly dependent on the initial fitting value of the input photon number. In Fig. 3(b), the error bars show apparent differences around their average measurement results when using a range of initial fitting values with ± 2% variations, resulting in a deterioration in the measurement accuracy. In Fig. 3(c), all curves are fitted with the same initial setting values, to avoid the error introduced by the fitting process in long time duration measurements.

Before solving this problem, we test the accuracy of the current system using another birefringent sample, a multi-order QWP at 808 nm. Similar to the organization in the measured HWP, Fig. 3(d), 3(e), and 3(f) show the phase retardation measurement with specified measurement bases, varying sample axes, and long test duration.

In Fig. 3(d), for a single measurement, the optimal fitting values of the phase retardations are 1.5556 and 1.5400 rad for H- and D-base, respectively. For a rotating sample, as shown in Fig. 3(e), the fitting phase retardations (left y-axis) with H-basis and D-basis are measured to be 1.5491 ± 0.0506 and 1.5365 ± 0.0408 rad, with maximum relative errors of 5.78% and 6.32%, respectively. In between, we remove some of the special points, such as $\theta$ =90° in H-base and $\theta$=45°/135° in D-base, because the output photon numbers have no relation with the phase retardation under these measurement bases, following the conditions in Eq.(5). Meanwhile, the time stability over 2.5 hours is shown in Fig. 3(f). The mean values of the phase retardation are 1.5568 ± 0.0075 rad and 1.5461 ± 0.0090 rad, with relative errors of -0.21% ± 0.48% and -0.89% ± 0.58%, respectively.

Similar to the discussion in HWP measurement, changing the initial fitting values of the incident photon counts within a range of ± 2% forms the error bar in Fig. 3(e). Unlike the error bar of the sample HWP, the phase retardation of the QWP shows little dependence on the default fitting value of the initial photon pair counts, giving more accurate and convincing results for the phase retardation measurement. All the above results indicate that our quantum ellipsometer is available for the phase retardation measurement of birefringent materials. To further enhance the accuracy of the current quantum ellipsometer, overcoming the dependence on the initial fitting conditions is the most critical step for the following discussion.

### 3.3 Comparison of Quantum and traditional ellipsometers with a compensator

The above section shows that the phase retardation may have a strong correlation with the initial fitting conditions according to the sample we measured. Considering that testing a sample with unknown reference values is common in practice, we need to improve the quantum ellipsometer to measure arbitrary birefringent materials with high accuracy. Following the analysis in Eq. (3) and (4), here we add a QWP (shown in the dotted box of Fig. 1(b)) in the idler path after the sample to solve this problem.

In the following discussion, all the purple lines in Fig. 4 represent the measurement results in the compensator-applied quantum ellipsometer. The coincidence photon counts are recorded based on the H-basis with the rotation angles of the sample (HWP and QWP) and the compensator fixed at 45° and 0°, respectively. Similar to Fig. 3, Fig. 4(a) and 4(d) show the coincidence photon counts for the true zero-order HWP and multi-order QWP, and the phase retardations of the two samples are estimated to be 3.1551 rad and 1.5680 rad. In Fig. 4(b) and 4(e), we fit the output curves with different sample axes ranging from 0° to 180°. The phase retardation (left y-axis) for the HWP and QWP samples are measured as 3.1403 ± 0.0633 rad and 1.5533 ± 0.0333 rad, while the maximum relative errors (right y-axis) between the standards are calculated as 3.6% and 4.2%.

Fig. 4(c) and 4(f) similarly show the time stability over 2.5 hours. The mean phase retardations (left y-axis) are 3.1463 ± 0.0060 and 1.5522 ± 0.0028 rad with average relative errors (right y-axis) of 0.39% ± 0.19% and -0.5% ± 0.18%, respectively.

With the introduction of compensator, the error bars in Fig. 4(b) and 4(e) apparently reveal that the measured phase retardations are independent of the initial fitting conditions for both the HWP and QWP, which significantly improves the accuracy of the phase retardation measurement in the current ellipsometer, especially for the HWP samples.

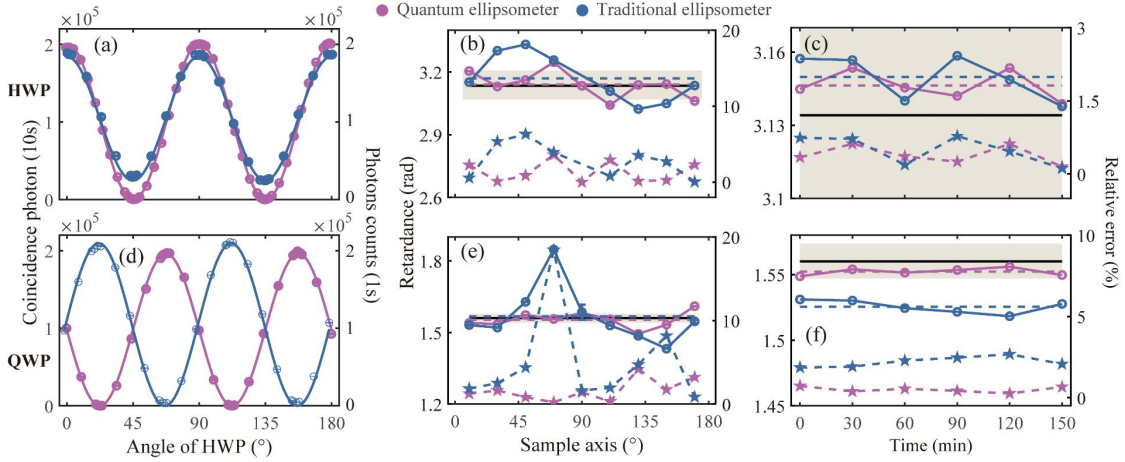

Fig. 4. Comparison of compensator-enabled quantum ellipsometer and traditional ellipsometer. (a) and (d): Single measurement for HWP and QWP. (b) and (e): The phase retardation with the varying sample axis ranging from 0° to 180°. (d) and (f): Time stability within 2.5 hours.

The above section shows the importance of introducing compensator in a quantum ellipsometer. In the following discussion, we also use the classical technique to measure the phase retardation of the sample with the same incident pump intensity. By setting the polarization of the pump photons horizontally, the input photons all pass through the counterclockwise path in the Sagnac loop, with the photons exiting the idler path with a particular polarization, transforming the dual-way quantum configuration into a classical one. In this way, the output counting rate is approximately $2.1 \times 10^5$ cps, and the dark count is 360 cps.

All the classical results are shown as the blue lines in Fig. 4. First, the phase retardations of the two samples are estimated to be 3.1571 and 1.5160 rad in a single measurement expressed by Fig. 4(a) and 4(d). Second, Fig. 4(b) and 4(e), similar to Fig. 3, show the retardance measured with different sample axes ranging from 0° to 180°. The average value and standard deviations are 3.1699 ± 0.1151 rad and 1.5676 ± 0.1187 rad, respectively. Third, the time stability of the classical ellipsometer system on the fixed axis is also measured within 2.5 hours in Fig. 4(c) and 4(f). The mean phase retardations are 3.1498 ± 0.0092 rad and 1.5255 ± 0.0050 rad, and the average relative errors are 0.5% ± 0.29% and -2.21% ± 0.32%, respectively. It appears that with the same pump power and designated sample axes, the quantum ellipsometer has improved accuracy and stability for phase retardation measurement compared with the classical system.

Here, in Table. 1, we list all the measurement results mentioned in the above sections. To evaluate the performance of the quantum ellipsometer with a QWP compensator, we compare the phase retardation of the samples measured by the quantum and classical ellipsometers with and without the QWP compensator. For long duration measurements, regardless of the use of a compensator, the quantum ellipsometer has a higher accuracy than the classical ellipsometer at the same pump intensity (cases 1 and 2 are better than case 3 in average values). This is due to the lower background noise and therefore higher signal-to-noise ratio of quantum ellipsometry, a benefit for finding the optical axes of the components and for fitting accuracy. The use of the compensator is reflected more in the improvement of the stability of the system for multiple measurements (cases 2 and 3 are better than case 1 in terms of standard deviation), because the introduction of the compensator directly improves the resolution for distinguishing small birefringence differences, while eliminating the dependence on the initial value for arbitrary sample conditions in the curve fitting process.

| Phase Retardance of target birefringent samples (rad) | | | |
|---|---|---|---|
| Ellipsometer configuration and test items | | HWP ($\delta_{std} = 3.1341 \pm 0.0628$) | QWP ($\delta_{std} = 1.56 \pm 0.0126$) |
| Case1: Quantum ellipsometer without QWP | Long duration measurement | 3.1131 ± 0.0321 (−0.67% ± 1.02%) | 1.5568 ± 0.0075 (−0.21% ± 0.48%) |
| | Varing sample axes | 3.0832 ± 0.0703 (−1.62% ± 2.24%) | 1.5491 ± 0.0506 (−0.70% ± 3.24%) |
| | Initial condition dependence | Yes | No |
| Case2: Quantum ellipsometer with QWP | Long duration measurement | 3.1463 ± 0.0060 (0.39% ± 0.19%) | 1.5522 ± 0.0028 (−0.50% ± 0.18%) |
| | Varing sample axes | 3.1403 ± 0.0633 (0.2% ± 2.02%) | 1.5533 ± 0.0333 (0.43% ± 2.13%) |
| | Initial condition dependence | No | No |
| Case3: Classical ellipsometer with QWP | Long duration measurement | 3.1498 ± 0.0092 (0.5% ± 0.29%) | 1.5255 ± 0.0050 (−2.21% ± 0.32%) |
| | Varing sample axes | 3.1699 ± 0.1151 (1.14% ± 3.67%) | 1.5676 ± 0.1187 (0.49% ± 7.67%) |
| | Initial condition dependence | No | No |

Table 1. Measurement results summary for the quantum and classical ellipsometer.

In addition, the changing sample axes tests compare three cases from another angle. Note that with this type of test, small changes in the incident point over the sample surface may occur when the sample itself is manually rotated. Since the birefringence distribution of the sample is not completely homogeneous, even for the commercial waveplate in our experiment, poorer accuracy and stability are shown in all three cases compared with the case where the sample is stationary. However, even in this situation, the quantum ellipsometer with the compensator in the middle row still shows optimal accuracy and stability compared to the other two cases, demonstrating the superior properties of our quantum ellipsometer system.

## 4. Conclusion

In summary, we use a highly qualified polarization-entangled photon source to establish a dual-way quantum ellipsometry system. Taking the commercially available HWP and QWP as samples, the experiment shows that the retardances are measured to be 3.1463 ± 0.0060 rad and 1.5522 ± 0.0028 rad under fixed measurement bases, with the long-term measurement deviations within 0.4% and 0.5%, respectively. In addition, the retardance under the varying optical axis of the sample is 3.1403 ± 0.0633 rad and 1.5533 ± 0.0333 rad, respectively, while the deviations from the standard values are all within 3%. The data show that the accuracy of the current quantum ellipsometer system can reach the nanometer scale, which is equivalent to the accuracy of the classical Senarmont method.

The phase retardations of the target samples measured by our system show better accuracy and stability in comparison with classical ellipsometry at the same incident photon intensity. The reason is supposed to lie in the configuration of the quantum ellipsometer. Different from the classical system, we replace the original one-way structure with a dual-way structure by separating the signal and idler entangled photon pairs into two different paths. Therefore, by collecting the coincidence counts of photon pairs instead of the output intensity in the one-way structure, the background noise is efficiently shielded so that the signal-to-noise ratio is improved. The reduced noise level significantly improves the accuracy of finding optical component axis and the visibility of the output intensity variation.

In addition, the quantum ellipsometer also shows its potential to open up a wide range of unique applications. Due to the dual-way configuration, the polarization measurement of the signal photon would determine the input polarization to the samples remotely on the idler photon side, allowing remote control in the retardance measurement. Furthermore, due to the extremely low intensity of the entangled photon sources compared to classical laser or white light sources, this quantum ellipsometer has the potential to detect photosensitive materials or fragile bioprocesses without the risk of photobleaching and thermal damage.

## Data availability statement

No new data were created or analysed in this study.

## Acknowledgement

We gratefully acknowledge the financial support from the National Key Research and Development Program of China (2022YFB3607700, 2022YFB3903102), National Natural Science Foundation of China (NSFC) (11934013, 92065101, 62005068), and Innovation Program for Quantum Science and Technology (2021ZD0301100), and the Space Debris Research Project of China (No. KJSP2020020202).

## Appendix A. Theoretical comparison between classical and quantum ellipsometer

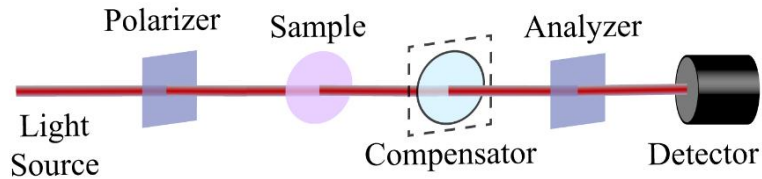

Fig.A1. Configuration of the Polarizer-Sample-Analyzer (PSA)

The differences between the classical and quantum ellipsometers can be seen from the above Fig.A1 and Fig.1(a) in the main text.

In the classical ellipsometer, each linear optical component inside the ellipsometer can be expressed by the Jones matrix, therefore the output polarization state described by the Jones vector $L_{out}$ is

$$L_{out} = AR(\mathrm{a})SR(-p)PL_{in}. \tag{A1}$$

Here A and P are the Jones matrices of the analyzer and polarizer, respectively. $R(\alpha) = \begin{pmatrix} \cos\alpha & \sin\alpha \\ -\sin\alpha & \cos\alpha \end{pmatrix}$ is the rotation matrix with angle $\alpha$ being the rotation angle. Considering the transmission rate in the sample to be polarization independent, the Jones matrix of the sample, S, can be expressed as

$$S = R(-\theta)S_0 R(\theta) = \begin{pmatrix} \cos\theta & -\sin\theta \\ \sin\theta & \cos\theta \end{pmatrix}\begin{pmatrix} 1 & 0 \\ 0 & e^{i\delta} \end{pmatrix}\begin{pmatrix} \cos\theta & \sin\theta \\ -\sin\theta & \cos\theta \end{pmatrix}. \tag{A.2}$$

By substituting all the Jones matrices into Eq. (A.1), we can derive the expression for the output light intensity in classical PSA-type ellipsometers:

$$I_{out1} = |E_{out}|^2 = \frac{I_0}{4}\left[2 + (1+\cos\delta)\cos(2(a-p)) + (1-\cos\delta)\cos(2(a+p-2\theta))\right]. \tag{A.3}$$

For the quantum ellipsometer in Fig.1(b), the polarization-entangled photon source is used as the light source of the transmission-type ellipsometry system, which is separated by a dual-way structure named signal and idler paths. Before entering the analyzer, the evolution of the entangled polarization state can be described as a tensor product of two unitary operators, namely $J_S \otimes J_I = S \otimes I$, showing that the idler photons pass through the sample while the signal photons remain invariant. Therefore, the two-photon state after passing through the sample is expressed in Eq. (1) in the main text.

To be consistent with the linear polarization transformation in Fig.A1, the following polarization analysis part of the quantum ellipsometer includes an HWP and a PBS on both sides. By rotating the HWP on either the signal side or the idler side, we can select a joint measurement base that is suitable for the sample birefringence measurement. Projective operators $\widehat{E}_m^{s(i)}$ and $\widehat{E}_n^{s(i)}$ with the HWP angle set at $h_{s(i)}$ in signal (idler) path are expressed as

$$\widehat{E}_m^{s(i)} = |M\rangle_{s(i)}\langle M| = \begin{pmatrix} \cos^2 2h_{s(i)} & \cos 2h_{s(i)}\sin 2h_{s(i)} \\ \cos 2h_{s(i)}\sin 2h_{s(i)} & \sin^2 2h_{s(i)} \end{pmatrix},$$

$$\widehat{E}_n^{s(i)} = |N\rangle_{s(i)}\langle N| = \begin{pmatrix} \sin^2 2h_{s(i)} & -\cos 2h_{s(i)}\sin 2h_{s(i)} \\ -\cos 2h_{s(i)}\sin 2h_{s(i)} & \cos^2 2h_{s(i)} \end{pmatrix}. \tag{A.4}$$

Therefore, the coincidence count under arbitrary joint projective measurement $\widehat{E}_m^{s(i)} \otimes \widehat{E}_n^{s(i)}$ is:

$$I_{out2} = I_0 \left| tr(\widehat{E}_m^{s(i)} \otimes \widehat{E}_n^{s(i)})|\phi_{sample}^+\rangle\langle\phi_{sample}^+| \right|^2$$

$$= \frac{I_0}{4}\left[2 - (1+\cos\delta)\cos(4(h_s + h_i)) - (1-\cos\delta)\cos(4(h_s - h_i - \theta))\right]. \tag{A.5}$$

Comparing Eq. (A.3) and Eq. (A.5), both the classical ellipsometer and the quantum ellipsometer proposed here have similar expressions. If we make a variable substitution $2h_i = a$ and $2h_s = \frac{\pi}{2} - p$, these two equations turn out to be the same. Such equivalence lies in the fact that the rotation of the HWP causes polarization changes twice as much as its rotation angle. At the same time, the polarizer and analyzer act on the photons with the same polarization while two HWPs are applied to a pair of entangled photons with reversed polarization, which is the reason why this phenomenon exists. The above discussion shows that the polarizer P and analyzer A in Supplement Fig. A1 play the same role as the HWPs in the signal and idler paths. Therefore, the concept of the quantum ellipsometer should be plausible by fixing one of the HWPs while rotating the other, similar to what PSA does.

The difference lies in the unique dual-way structure of the quantum ellipsometer, which completely separates the "control" and "measurement" terminals, allowing remote control of the sample birefringence measurement. In addition, unlike the classical ellipsometer, both the base selection and the measurement can be performed without disturbing the light and sample interaction process, enhancing the stability of the operating process.